\documentstyle[12pt,aps,epsf]{revtex} 
\begin{document}
\draft
\def\half{{1\over 2}} 
\def\vopo{(VO)$_2$P$_2$O$_7$\ } 
\def\vodpo{VODPO$_4 \cdot 1/2$D$_2$O\ } 
\def\vohpo{VOHPO$_4 \cdot  1/2$H$_2$O\ } 
\newcommand{\lapprox}{\stackrel{<}{\scriptstyle \sim}} 
\newcommand{\gapprox}{\stackrel{>}{\scriptstyle \sim}}

\date{\today}
\setlength{\footheight}{0.cm}
\setlength{\textwidth}{16.0cm}
\setlength{\textheight}{23.0cm}
\setlength{\fboxsep}{2mm} 
\pagestyle{plain}
\def\boxit#1{\leavevmode\hbox{\vrule\vtop
    {\vbox{\hrule\kern1pt\hbox{\kern1pt
    \strut#1\kern1pt}}\kern1pt\hrule}\vrule}}

\title{\bf Excitations and Possible Bound States in the $S=1/2$ 
Alternating Chain Compound \vopo}

\author{
D.A. Tennant$^1$, S.E. Nagler$^1$, T. Barnes$^{1,2}, $ A.W. Garrett$^3$,
J. Riera$^4$,
and B.C. Sales$^1$ }

\address{
$^1$Oak Ridge National Laboratory, Oak Ridge, TN 37831-6393, USA  \\  
$^2$University of Tennessee, 
Knoxville, TN 37996-1501, USA \\
$^3$University of Florida,
Gainesville, FL 32611-0448, USA\\
$^4$Insituto de Fisica Rosario, 2000 Rosario, Argentina\\
}

\maketitle

\begin{center}
{\bf Abstract}
\end{center}

\begin{abstract}
Magnetic excitations in an array of \vopo single crystals have been 
measured using 
inelastic neutron scattering. Until now, \vopo has been thought 
of as a two-leg 
antiferromagnetic Heisenberg spin ladder with chains running in 
the $a$-direction.  
The present results show unequivocally that \vopo is best 
described as an alternating 
spin-chain directed along the crystallographic $b$-direction. 
In addition to the expected magnon with magnetic zone-center 
energy gap $\Delta = 3.1$ meV, 
a second excitation is observed at an energy just below $2\Delta$.  
The higher mode may be a triplet two-magnon bound state. Numerical results
in support of bound modes are presented. 
\end{abstract}

{\bf Corresponding Author: } \\
Dr Alan Tennant\\
Bldg 7962 MS 6393, Solid State Division \\
Oak Ridge National Laboratory, Oak Ridge, Tennessee 37831-6393,
U.S.A. \\
tel.: +1-423-576-7747 \\
fax.: +1-423-574-6268 \\
email: alan@phonon.ssd.ornl.gov \\

\newpage

The S=1/2 alternating Heisenberg chain (AHC) is a fascinating
quantum system that is currently the subject of much interest.  We have
established \cite{gntsb}
that the material \vopo, previously considered to be
a spin ladder, is in fact an excellent realization of the AHC.  The physics
of the AHC is
also very relevant to spin-Peierls materials such as CuGeO3 \cite{nishi}.  
Recent theoretical work \cite{uhrig}
on the AHC underscored the potential importance of two-magnon bound modes.
In this paper, we
review our neutron scattering experiments on the alternating chain material
\vopo.
In addition to the expected one-magnon excitations, we observe an extra
mode, which may be
a two-magnon bound state.  We follow with a discussion of some related
theoretical issues.

The crystal structure of \vopo is nearly 
orthorhombic, with a slight monoclinic distortion so that the 
space group is P2$_{1}$\cite{nhs}. 
The room temperature lattice parameters 
are $a$=7.73\AA, $b$=16.59\AA, $c$=9.58\AA
and $\beta$=89.98$^{\circ}$.
The magnetic properties of \vopo arise from $S=1/2$ V$^{4+}$ 
ions situated within 
distorted VO$_6$ octahedra. 
Face-sharing pairs of VO$_6$ octahedra are stacked in two-leg 
ladder structures 
oriented along the $a$-axis. 
The ladders are separated by large, 
covalently bonded PO$_4$ complexes.  
The structure is illustrated
schematically in figure 1.

The susceptibility of \vopo powder \cite{jjgj}
can be accurately reproduced by either a spin 
ladder (with $J_\| \approx J_\perp$) or by an alternating 
chain \cite{jjgj,br}, but the 
expectation that the PO$_4$ group would provide a weak 
superexchange path led to a general acceptance
of the spin ladder interpretation of \vopo.   
Pulsed inelastic neutron scattering
measurements on \vopo powders \cite{ebbj} showed a spin gap 
of 3.7 meV, which was interpreted as further support 
for the ladder model.

Beltr\'an-Porter {\it et al.}\cite{euro_chemists} examined the
superexchange pathways in several vanadyl phosphate compounds, 
and were led to question the spin-ladder interpretation of \vopo. 
Instead they proposed that an alternating V-O-V-PO$_4$-V chain
in the $b$-direction was a more likely magnetic system. 
The observation of a second spin excitation 
near 6 meV (not predicted by the ladder model) 
in a recent triple-axis neutron scattering 
experiment \cite{gnbs} on \vopo powder, and
the discovery of strong superexchange through PO$_4$
groups in the precursor compound \vodpo \cite{vodpo_paper}, 
also cast doubt on the spin ladder interpretation. 
For these reasons we undertook studies of the spin 
dynamics in \vopo single crystals.  

Measurements of the excitations were made using a
single crystal array of approx. 200 oriented 
\vopo crystals of typical size 1x1x0.25 mm$^3$.
The resulting sample had an effective 
mosaic spread of  $8-10^\circ$ FWHM.  
Inelastic neutron scattering measurements 
were carried out using 
triple-axis spectrometers at the HFIR reactor, 
Oak Ridge National Laboratory; full 
experimental details can be found in \cite{gntsb}.

Scans \cite{gntsb} at $T=10$K showed two modes of 
roughly equal strength at the antiferromagnetic zone-centre
$(0,\pi,0)$ at energies of $\Delta_{l}=3.12(3)$ meV and 
$\Delta_{u}=5.75(2)$ meV. 
Full resolution convolutions with the fitted dispersion
showed these modes to be resolution limited.
The disappearance of both modes at higher temperatures
confirmed their magnetic origin. The modes were found to 
track approximately in $Q$ close to $(0,\pi,0)$ (see Figure (2)).
At the zone-boundary $(0,\pi/2,0)$ only a single mode was
observed at an energy of $\approx 15$meV. 
Because of poor instrumental resolution
it was not possible to tell whether the modes had coalesced or
were simply not resolved.  

Fig. 2 shows the measured dispersion for both modes
along $a^*,b^*$ and $c^*$. 
The excitation energy is almost 
independent of $Q_{c}$ 
(middle panel), implying a very weak coupling along $c$.  
The dependence of energy on $Q_{a}$ is much 
weaker than on $Q_{b}$ and is ferromagnetic.
The strong $Q_{b}$ dependence implies that the 
exchange coupling is dominantly 
one-dimensional along the $b$-direction, 
confirming the V-O-V-PO$_4$-V alternating chain
proposed in \cite{euro_chemists} and \cite{vodpo_paper}. 

For any exchange alternation $-$ as occurs
with the two inequivalent exchanges along 
$b$ $-$ a gap should appear in the dispersion
(as observed) and the absence of magnetic ordering in \vopo
is consistent with a singlet ground state. However the observation of
an extra mode requires a more thorough theoretical investigation.   
Since the high temperature limit
of the magnetic susceptibility\cite{jjgj} is consistent
with expectations for 
simple $S=1/2$, $g=2$ spins, the 
possibility that the upper mode is an
additional  low lying single ion 
excitation can be ruled out.
Two other plausible explanations of the second
peak are (a) splitting
due to an exchange anisotropy, and
(b) a physical two-magnon bound state.

Although a pseudo-Boson calculation including exchange
anisotropy gave an excellent fit to the dispersion \cite{gntsb}
$-$ the solid line in Figure (2) is a fit 
to this model \cite{gntsb} $-$ considerable 
exchange anisotropy ($\approx 15 \%$)
was necessary to account for the mode 
splitting. Recent single crystal magnetic susceptibility 
measurements \cite{thompson} were quantitatively consistent with the previous 
powder results\cite{jjgj} and found little if any evidence for anisotropy.
Also the coupling in the precursor compound \vodpo was found to be 
consistent with isotropic exchange \cite{vodpo_paper} suggesting that one should 
seek another explanation for the second mode, and
because the energy of
the upper mode at $(0,\pi,0)$, $\Delta_{u}$, is 
just below $2\Delta_{l}$, a bound two-magnon mode
may provide a good explanation.
In support of this explanation,
preliminary high-field measurements show splitting 
of both modes which is consistent with both modes being triplets
\cite{vopo_field}.

To gain some insight into the formation of
two-magnon modes in \vopo we have studied the
$S=1/2$ AHC using numerical techniques.
The isotropic AHC Hamiltonian is
\begin{equation}
H =
 \sum_{i=1}^{L/2}
\ J \; {\vec S_{2i-1}} \cdot {\vec S_{2i}}
+ \alpha J \;  {\vec S_{2i}} \cdot {\vec S_{2i+1}} \ ,
\end{equation}
where $J>0$ and $1 \ge \alpha \ge 0$.
Equation (1) has been studied analytically and numerically 
over many years, but it had not been appreciated until recently
that $S=0$ and $S=1$ bound magnon states may form.
Uhrig and Schulz \cite{uhrig} have used field theory and RPA methods
to study these modes at $k=\pi/2$ and $k=0,\pi$.
The existence of these bound states depends subtly on the 
kinetic and potential energies of pair formation, and
occur for only certain values of $k$.

Perturbation theory in 
$\alpha$ about the dimer limit ($\alpha=0$) 
provides a quantitative basis for understanding the 
excitations for small $\alpha$, and also provides
insight into the competition between potential and 
kinetic energy effects in bound states \cite{brt}. 
Figure (3) shows the one- and
two-magnon excitation spectra calculated within a 
simplified approximate first order 
(one- and two-magnon manifold) treatment of the AHC.
At $k=\pi/2$ there is a node in the two-magnon continuum
which corresponds to a degeneracy in the 
total kinetic energy $\omega(k_{1})+\omega(\pi/2-k_{1})$ 
of two magnons. The $S=0$ and $S=1$ 
bound states lie well below the continuum lower boundary.
However at $k=0$ and $\pi$ only the $S=0$ bound state is seen. 
The continuum is much broader at $k=0,\pi$ indicating 
larger mixing effects which disrupt the $S=1$ 
bound state. Although no $S=1$ bound state forms, the
attractive potential still leads to a strongly enhanced 
scattering cross-section $S(Q,\omega)$ at the 
continuum lower boundary \cite{brt}, see dashed line 
in Figure (4). The $S=1$ bound state appears 
clearly at the $k=\pi/2$ point (solid line in Figure (4)). 
It should be noted
that the neutron scattering cross-section for the
$S=0$ mode is zero, however this mode may be visible 
by light scattering \cite{brt}.

Harris \cite{abh} used a reciprocal space
perturbation theory to calculate the ground state and excited
state energy up to $O(\alpha^3)$. This gives a
$k=0,\pi$ energy gap of
$E_{gap}=J(1-\alpha/2-3 \alpha^2 /8 + \alpha^3 /32)$.
However these results can
be derived more easily in real space \cite{brt}, and
in the case of the ground state energy, we have extended the 
calculation to $O(\alpha^5)$,
\begin{eqnarray}
e_0(\alpha)/J=-3/2^3-(3/2^6) \cdot \alpha^2 - (3/2^8) \cdot \alpha^3
\nonumber \\
- (13/2^{12}) \cdot \alpha^4 - (95/3) \cdot (1/2^{14}) \cdot 
\alpha^5 - O(\alpha^6) .
\end{eqnarray}
The perturbation series appears to be rapidly converging for 
$\alpha \le 0.5$, and may have a radius of convergence
of unity.

Because \vopo has $\alpha \approx 0.8$ \cite{gntsb}, we have 
used a numerical Lanczos algorithm on finite $L=4n$ lattices of up to $L=28$ 
and with approximately 14 place accuracy to study the
ground states and binding energies up to similar values of $\alpha$.
Full details will be given elsewhere \cite{brt}.
Figure (4) shows the calculated binding energies of the $S=0$  
bound mode at $k=\pi/2$, and $k=0$, as well as
those for the $S=1$ bound mode at $k=\pi/2$.
The results show strong binding at $\alpha=0.8$ 
of the $S=0$ mode at $k=\pi/2$ but the situation is
not clear for $k=0$. They also suggest weak binding for 
the S=1 mode at $\pi/2$ at the alternation for \vopo. 
Unfortunately finite size effects precluded an accurate determination 
of this binding energy.

In order to make a quantitative comparison with
\vopo $S(Q,\omega)$ is required for the bound modes
and continuum. We are currently undertaking calculations
to quantify this. 
The effects of interchain coupling have been neglected and 
these may enhance the binding. Next-nearest 
neighour exchange within the chains may have a similar
effect. The $\alpha$ perturbation theory provides a
useful quantitative guide to such effects, and
further theoretical studies are in progress.
We also note that similar dynamics are also 
important in many other low-dimensional Hamiltonians
such as spin ladders and we shall present some work on those
in the future.

In conclusion, we have measured an extra mode in the
alternating chain system \vopo. The evidence suggests
that this is a two-magnon bound state. Perturbation theory 
and Lanczos calculations give an insight into the formation
of bound modes.

We thank J.Thompson
for sharing his susceptibility results with us prior to publication.
Oak Ridge National Laboratory is 
managed for the U.S. D.O.E.
by Lockheed Martin Energy Research Corporation under 
contract DE-AC05-96OR22464. Work at U.F. is 
supported by the U.S. D.O.E. under contract DE-FG05-96ER45280.

\newpage

\begin{center}
{\Large Figure Captions}
\end{center}

\begin{figure}
{Figure~1.
Schematic depiction of the structure and magnetic interactions in VOPO.  
The spin ladder model previously thought to describe VOPO has 
nearest neighbor exchange constants $J_{\parallel}$ along 
the $a$ (``ladder'') direction and $J_{\perp}$ along 
the $b$ (``rung'') direction.  In the alternating chain model, 
nearest neighbor V$^{4+}$ ions are alternately coupled by 
constants $J_{1}$ and $J_{2}$ along the $b$ (chain) direction.  
Neighboring spins in adjacent chains are coupled by $J_{a}$.  
Magnetic coupling in the $c$ direction is negligible.
}
\end{figure}

\begin{figure}
{Figure~2.
Measured dispersion of magnetic excitations in VOPO at T = 10K.  
When not visible error bars are smaller than the size of the plotted symbols.  
Filled circles (open diamonds) are points from the lower (upper) energy mode.
The solid lines are dispersion curves calculated using 
parameters obtained by fitting to a pseudo-Boson model \cite{gntsb}. 
Wavevectors are plotted in units corresponding 
to the VOPO reciprocal lattice.
}
\end{figure}

\begin{figure}
{Figure~3.
Schematic depiction of the one- and two-magnon excitation spectra
of the $S=1/2$ AHC with an alternation of $\alpha=0.2$.
An $S=1$ bound mode appears below the continuum at $k \approx \pi/2$.
The more deeply bound $S=0$ mode (dashed line) is not visible to 
neutrons scattering.
}
\end{figure}

\begin{figure}
{Figure~4.
Calculated $S(Q,\omega)$ for constant-Q scans at
$k=\pi/2$ (solid line) and $k=\pi$ (dashed line)
using the first order perturbation approach with
$\alpha=0.2$.
}
\end{figure}

\begin{figure}
{Figure~5.
Calculated binding energies of the $S=0$ and $S=1$ 
bound states using a Lanczos method \cite{brt} at 
$k=0$ and $\pi/2$. The binding energies are given in units
of $J$.
}
\end{figure}

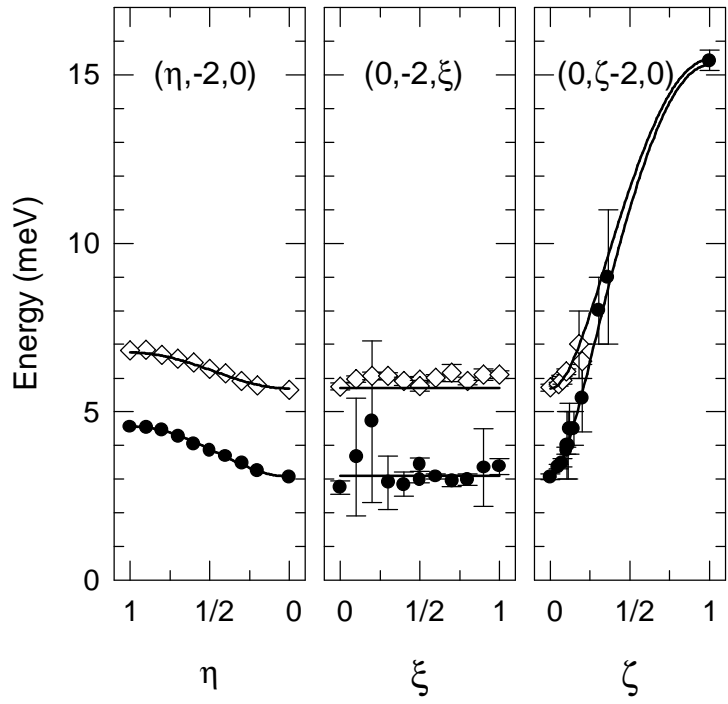
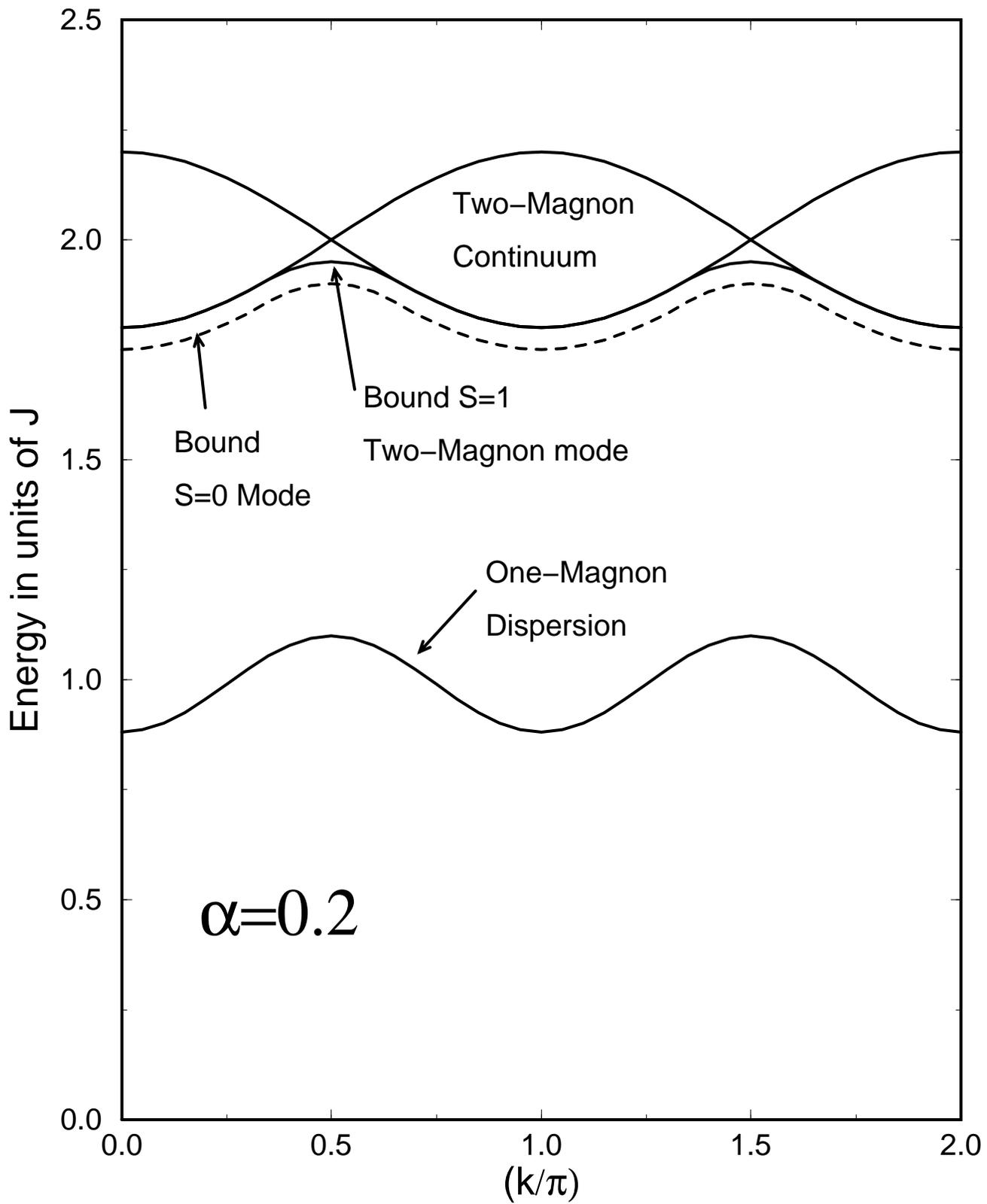
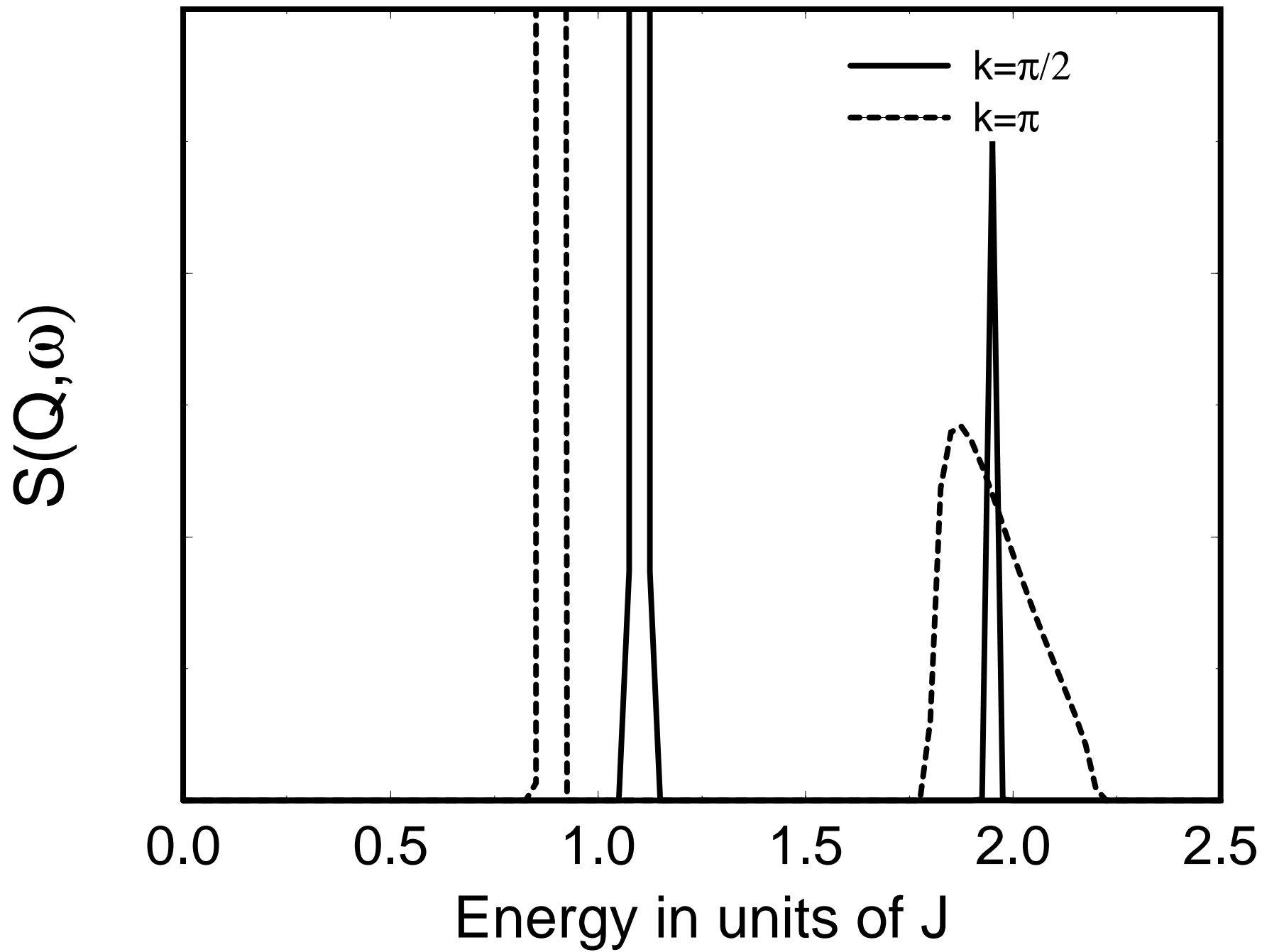
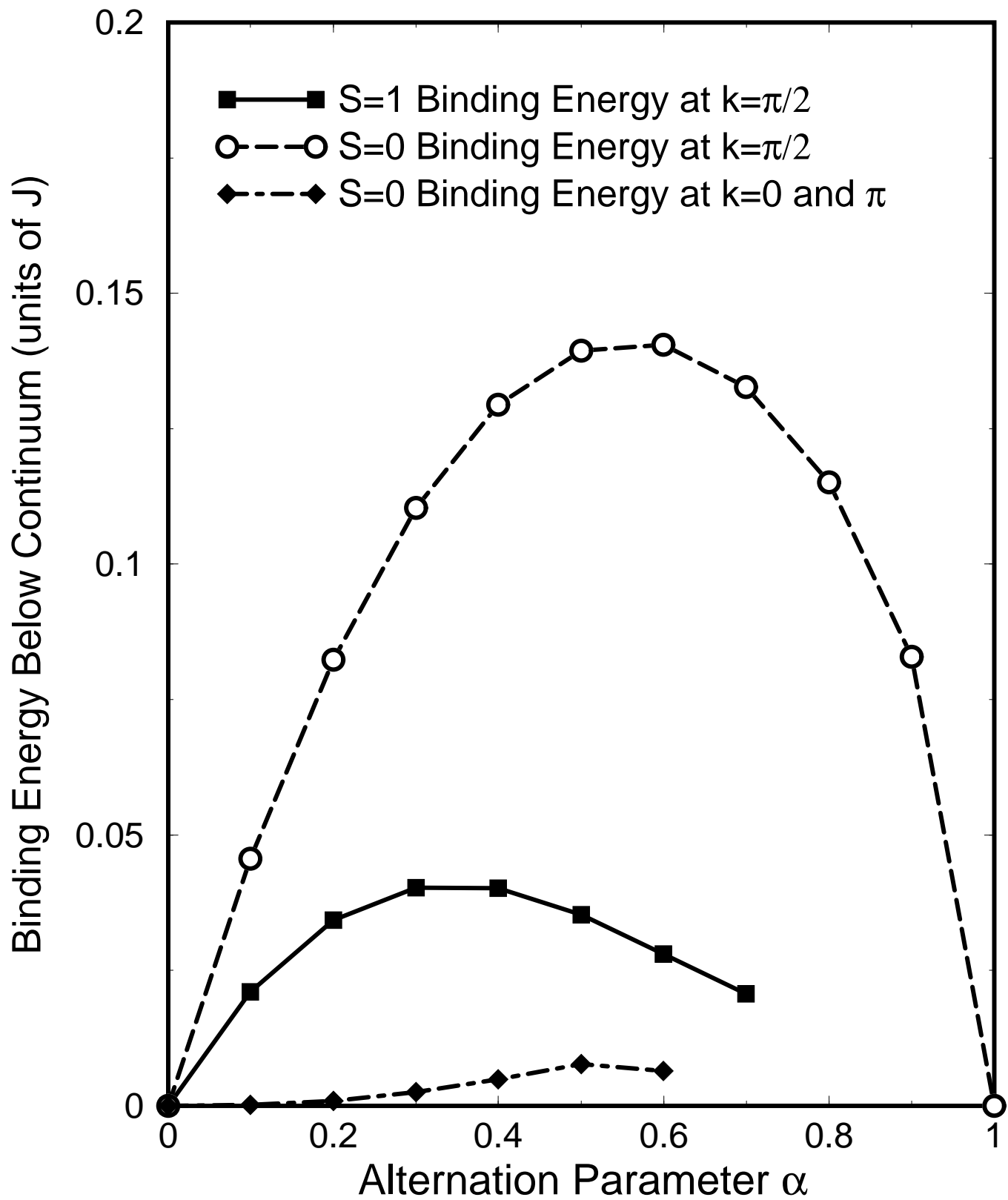
\end{document}